\documentclass[a4paper,9 pt, conference]{ieeeconf}
\IEEEoverridecommandlockouts 
\usepackage{amsmath,amssymb,latexsym}
\usepackage{SDC_macros_noalgo}
\usepackage[dvips]{graphicx}
\usepackage{graphicx}
\usepackage{psfrag}
\usepackage{epsfig}
\usepackage{float}
\usepackage{array}
\usepackage{mathrsfs}
\usepackage[geometry]{ifsym}
\usepackage{color}
\usepackage{fancyhdr}
\usepackage{algorithmic}
\usepackage{cite}

\title{\bf Extremum Seeking-based Iterative Learning Linear MPC}
\author{ Mouhacine Benosman, Stefano Di Cairano, Avishai Weiss
\thanks{M. Benosman (m{\_}benosman@ieee.org), S. Di Cairano (dicairano@ieee.org), and A. Weiss (weiss@merl.com)  are with Mitsubishi Electric Research Laboratories,
201 Broadway Street, Cambridge, MA 02139, USA. Accepted at the
IEEE MSC 2014}
 }

\date{}

\begin{document}
%\title{\bf Lyapunov-Based Control of Sway for Elevators' Ropes}
%\author{M. Benosman*\\
%  * Mitsubishi Electric Research Laboratories,\\
%  201 Broadway Street, Cambridge, MA 02139, USA\\E-mail:
%benosman@merl.com\\\\ \copyright MERL
%\today\\{\color{red}{CONFIDENTIAL}}}
%\date{}
\maketitle\thispagestyle{empty}
%\tableofcontents
%\newpage
\begin{abstract}
In this work we study the problem of adaptive MPC for linear
time-invariant uncertain models. We assume linear models with
parametric uncertainties, and propose an iterative multi-variable
extremum seeking (MES)-based learning MPC algorithm to learn
on-line the uncertain parameters and update the MPC model. We show
the effectiveness of this algorithm on a DC servo motor control
example.
\end{abstract}
\section{Introduction}
Model predictive control (MPC) \cite{Mayne2000} is a model-based
framework for optimal control of
 constrained multi-variable systems. MPC is based on the repeated,
receding horizon solution of a finite-time optimal control problem
formulated from the system dynamics, constraints on system states,
inputs, outputs, and a cost function describing the control
objective. MPC has been applied to several applications such as
aerospace \cite{hartley12},\cite{DPK12}, automotive
\cite{DYBKH11},\cite{DTBB12}, and mechatronic systems
\cite{DBKH07b},\cite{grancharova09}. Since MPC is a model-based
controller, its performance inevitably depends on the quality of
the prediction model used in the optimal control computation.

% Different classes of MPC have been developed, differentiated by the properties of the system model, cost function, and constraints. In this paper we focus on linear-quadratic model predictive control for regulation and tracking, i.e. linear model predictive control.

In contrast, extremum seeking (ES) control is a well known
approach where the extremum of a cost function associated with a
given process performance (under some conditions) is found without
the need for detailed modeling information, see, e.g.,
\cite{ariyur2003real,AK02,Nes09}. Several ES algorithms (and
associated stability analyses) have been proposed,
\cite{k00,AK02,TNM06,Nes09,TNM06,ariyur2003real,rotea2000analysis,GDD13},
and many applications of ES have been reported
\cite{ZGD03,HGPD08,ZO12,benosmanatincacc13,benosmanatincecc13}.

The idea that we introduce in this work, is that the performance of
a model-based MPC controller can be combined with the robustness
of a model-free ES learning algorithm for simultaneous
identification and control of linear time-invariant systems with
structural uncertainties. While regulation and identification are
seemingly conflicting objectives, by identifying (or
re-identifying) the system dynamics online and updating the MPC
prediction model, the closed-loop performance may be enhanced relative
to a standard MPC scheme that uses an inaccurate (or outdated)
model. The optimal solution to this trade-off between
identification and control is given by a dynamic program
\cite{feldbaum1960dual}, which, for many applications, is
computationally intractable. As a result, many suboptimal
techniques and heuristics have been developed in recent
years -- often in a \emph{receding horizon} framework which is suitable for integration with MPC.
%over the entire optimization horizon

In \cite{Sousa1999}, an approximation of the dynamic program is
developed for a linear input-output map with no dynamics.
Approaches for more complex systems avoid dynamic programming
altogether and, instead, sub-optimally trade off between inputs that
excite the system and inputs that regulate the state. Excitation
signals are often designed to satisfy \emph{persistency of
excitation} conditions. For example, a dithering signal may be
added on top of the nominal control \cite{Sotomayor2009551},
although difficulties arise in determining the amplitude of the
signal, and the dither indiscriminately adds noise to the process.
More sophisticated schemes employ \emph{optimal input design},
usually in the frequency domain, where maximizing the Fisher
information matrix can be cast as a semi-definite program
\cite{jansson2005input}. However, design in the frequency domain
leads to difficulties with constraints that are more naturally
addressed in the time domain, e.g., input (and possibly output) amplitude constraints.
While the problem formulation in the time domain is highly
non-convex, developing such techniques is desirable, and thus the
focus of recent work
\cite{marafioti2013persistently,vzavcekova2013persistent,ACS:ACS2370,genceli1996new,larssonmo
del,heirung2013adaptive,adetola2009adaptive,gonzalez2013model,aswani2013provably}.

In this preliminary work, we aim at proposing an alternative
approach to realize an iterative learning-based adaptive MPC. We
introduce an approach for a multi-variable extremum seeking
(MES)-based iterative learning MPC that merges a model-based
linear MPC algorithm with a model-free MES algorithm to realize an
iterative learning MPC that adjusts to structured model
uncertainties. This approach is an extension to the recent MPC
framework reported in~\cite{B014}, where the author proposed to
use MES with model-based nonlinear control to design
learning-based adaptive controllers for a class of nonlinear
systems.

The paper is organized as follows. We start
the paper with some preliminaries in Section \ref{section0}. In
Section \ref{section2}, we first recall the nominal MPC algorithm
and then present the main result of the paper, namely, the
iterative learning MPC, with
  a discussion of the algorithm stability.
  Section \ref{section3} is dedicated to a DC servo-motor case study. Finally, we conclude the paper with a brief summary of the results in Section
 \ref{section4}.
 \section{Notations and preliminaries}\label{section0}
Throughout the paper, $\mathbb{R}$, $\mathbb{Z}_{0+}$,
$\mathbb{Z}_{[i,j]}$ denote the set of real numbers, positive
integers, and positive integers from $i$ to $j$, respectively. For
$x\in \mathbb{R}^{N}$ we define
 $||x||=\sqrt{x^{T}x}$, we denote by $A_{ij},\;i=1,\ldots ,n,\;j=1,\ldots ,m$ the elements of the
matrix
 $A$, and by $[A]_{i}$ the vector equal to the line $i$ of the matrix $A$. We denote by $||A||_{2}$ the spectral matrix norm, and by $||x||_{A}=\sqrt{x^{T}Ax}$. We denote
 by $x(i|k)$ the value of $x$ at the time sample $i$
 and the MPC cycle $k$. In the sequel, when we use the term well-posed
 optimization problem, we mean that the problem admits a unique solution,
 which is a continuous function of the initial conditions
 \cite{J02}.
\section{Iterative learning-based adaptive MPC}\label{section2}

\subsection{Control objective}
We want to design an adaptive controller that solves regulation
and tracking problems for linear time-invariant systems with
structural model uncertainties under state, input, and output
constraints.\\ In what follows, we first present the nominal MPC
problem, i.e., without model uncertainties, and then extend this
nominal controller to its adaptive form by merging it with an MES
algorithm.
\subsection{Constrained linear nominal MPC}\label{nominal_mpc}

Consider a linear MPC, based on
the nominal linear prediction model
 \bsube\label{eq:linsys}
  \beqar
     x(k+1)&=&Ax(k)+Bu(k),\label{eq:linsys-state}\\
     y(k)&=&Cx(k)+Du(k),\label{eq:linsys-output}
     \eeqar
   \esube
where $x\in\rr^n$, $u\in\rr^m$, $y\in\rr^p$ are the state, input,
and output vectors  subject to constraints
\bsube\label{eq:linconstr}
 \beqar
           \xmin\leq x(k) \leq \xmax,   \\
          \umin\leq  u(k) \leq \umax, \\
         \ymin\leq  y(k) \leq \ymax,
            \eeqar
  \esube
where  $\xmin,\xmax\in\rr^n$, $\umin,\umax\in\rr^m$, and
$\ymin,\ymax\in\rr^p$ are the lower and upper bounds on the state,
input, and output vectors, respectively. At every control cycle
$k\in\zz_{0+}$, MPC solves the finite horizon
optimal control problem
  \bsube\label{eq:OCP}
  \beqar
\min_{U(k)}&&\sum_{i=0}^{N-1}\|x(i|k)\|^2_{Q_{\rm M}}+\|{u(i|k)\|^2_{R_{\rm
M}}}  \label{eq:OCP-cost}  \\ &&\quad+ \|x(N|k)\|^2_{P_{\rm
M}}, \nonumber
 \\%+\|y(i|k)\|_2^{S_{\rm  M}} \\
   {\rm s.t.}
         && x(i+1|k)=Ax(i|k)+Bu(i|k),\  \qquad
\label{eq:OCP-dynam}\\
         && y(i|k)=Cx(i|k)+Du(i|k),\
\label{eq:OCP-output}\\
        && \xmin\leq x(i|k) \leq \xmax,\  i\in\zz_{[1,N_c]},
\label{eq:OCP-Xcnstr} \\
       && \umin\leq  u(i|k) \leq \umax,\  i\in\zz_{[0,N_{cu}-1]},
\label{eq:OCP-Ucnstr} \\
       && \ymin\leq  y(i|k) \leq \ymax,\  i\in\zz_{[0,N_c]},
\label{eq:OCP-Ycnstr} \\
        &&  u(i|k)=K_f x(i|k),\  i\in\zz_{[N_u,N-1]},
\label{eq:OCP-tControl} \\
        && x(0|k) = x(k), \label{eq:OCP-initX}
  \eeqar
\esube where $Q_{\rm M}\geq 0$, $P_{\rm M},R_{\rm M}> 0$ are
symmetric weight matrices of appropriate dimensions, $N$ is the
prediction horizon, $N_u\leq N$ is the control horizon (the number
of free control moves), $N_{cu}\leq N$, $N_c\leq N-1$ are the input
and output constraint horizons along which the constraints are
enforced. The performance criterion  is
defined by~\eqref{eq:OCP-cost},
and~\eqref{eq:OCP-Xcnstr}--\eqref{eq:OCP-Ycnstr} enforce the
constraints. Equation~\eqref{eq:OCP-tControl} defines the
pre-assigned terminal controller where $K_f\in\rr^{m\times n}$, so
that the optimization vector effectively is $U(k)=[u'(0|k) \ldots
u'(N_u-1|k)]'\in\rr^{N_um}$.

Although the optimal control problem~\eqref{eq:OCP} does not
explicitly mention a reference, tracking is achieved by including
in the state update equation~\eqref{eq:linsys-state} the reference
prediction dynamics
\beq\label{eq:refModel}  {r_{r}(k+1)=A_rr_r(k),}\eeq
and an additional output in~\eqref{eq:linsys-output} representing
the tracking error
\begin{equation}\label{tracking_error}
{ y_e(k)=Cx(k)-C_rr_r(k),}
\end{equation}
 which is then accounted for in the cost
function \eqref{eq:OCP-cost} as  later shown in the example (see
also \cite{DTBB12} for an example in a real world application). At
time $k$, the MPC problem~\eqref{eq:OCP} is initialized with the
current state value $x(k)$ by~\eqref{eq:OCP-initX} and solved to
obtain the optimal sequence $U^*(k)$. Then, the input
$u(k)=u_\MPC(k)=u^*(0|k)=[I_m\ 0\ \ldots \ 0]U(k)$ is applied to
the system.
\subsection{Learning-based adaptive MPC algorithm}
Consider now, the system (\ref{eq:linsys}), with structural
uncertainties, such that
 \bsube\label{eq:linsys_uncertain}
  \beqar
     x(k+1)&=&(A+\Delta A)x(k)+(B+\Delta B)u(k)\label{eq:linsys-state_uncertain}\\
     y(k)&=&(C+\Delta C)x(k)+(D+\Delta D)u(k),\label{eq:linsys-output_uncertain}
     \eeqar
   \esube
   with the following assumptions.
   \begin{assumption}\label{assumption1}
The constant uncertainty matrices $\Delta A,\;\Delta B,\;\Delta C$
and $\Delta D$, are bounded, s.t. $||\Delta A||_{2}\leq l_{A}$,
$||\Delta B||_{2}\leq l_{B}$, $||\Delta C||_{2}\leq l_{C}$,
$||\Delta D||_{2}\leq l_{D}$, with
$l_{A},\;l_{B},\;l_{C},\;l_{D}>0$.
   \end{assumption}
\begin{assumption}\label{assumption2}
There exists non empty convex sets
$\mathcal{K}_{a}\subset\mathbb{R}^{n\times n}$,
$\mathcal{K}_{b}\subset\mathbb{R}^{n\times m}$,
$\mathcal{K}_{c}\subset\mathbb{R}^{p\times n}$, and
$\mathcal{K}_{d}\subset\mathbb{R}^{p\times m}$, such that
$A+\Delta A\in\mathcal{K}_{a}$ for all $\Delta A$ such that $||\Delta A||_{2}\leq l_{A}$,
$B+\Delta B\in\mathcal{K}_{b}$ for all $\Delta B$ such that $||\Delta B||_{2}\leq l_{B}$,
$C+\Delta C\in\mathcal{K}_{c}$ for all $\Delta C$ such that $||\Delta C||_{2}\leq l_{C}$,
$D+\Delta D\in\mathcal{K}_{d}$ for all $\Delta D$ such that $||\Delta D||_{2}\leq l_{D}$,.
   \end{assumption}
\begin{assumption}\label{assumption3}
The iterative learning MPC problem (\ref{eq:OCP}) (and the associated
reference tracking extension), where we substitute the model with structural uncertainty
(\ref{eq:linsys_uncertain}) for the nominal model (\ref{eq:OCP-dynam}) and
(\ref{eq:OCP-output}), is a well-posed optimization problem for
any matrices $A+\Delta A\in\mathcal{K}_{a}$,
$B+\Delta B\in\mathcal{K}_{b}$,
$C+\Delta C\in\mathcal{K}_{c}$,
$D+\Delta D\in\mathcal{K}_{d}$.
   \end{assumption}

Under these assumptions, we postulate the following: If we solve
the iterative learning MPC problem (\ref{eq:OCP}), where we
substitute (\ref{eq:linsys_uncertain}) for (\ref{eq:OCP-dynam})
and (\ref{eq:OCP-output}), iteratively, such that, at each new
iteration we update our knowledge of the uncertain matrices
$\Delta A$, $\Delta B$, $\Delta C$, and $\Delta D$, using a
model-free learning algorithm, in our case the extremum seeking
algorithm, we claim that, if we can improve over the iterations
the MPC model, i.e., learn over iterations the uncertainties, then
we can improve over time the MPC performance, i.e., either in the
stabilization or in the tracking. Before formulating this idea in
terms of an algorithm, we briefly recall the principle of
model-free multi-variable extremum seeking (MES) control.

To use the MES learning algorithm, we define the cost function to
be minimized as
\begin{equation}\label{Q}
Q(\hat\Delta)=F(y_{e}(\hat\Delta)),
\end{equation}
where $\hat\Delta$ is the vector obtained by concatenating all the
elements of the estimated uncertainty matrices $\Delta\hat{ A}$,
$\Delta\hat{ B}$, $\Delta\hat{C}$ and $\Delta\hat{D}$,
$F:\;\mathbb{R}^{p}\rightarrow\mathbb{R},\;F(0)=0,\;F(y_{e})>0$
for $y_{e}\neq 0$.

%To use the MES learning algorithm, we define the cost function to
%be minimized as
%\begin{equation}\label{Q}
%Q(\hat\Delta)=\tilde F(\psi({y_{e}}^{[t-\delta T_{\rm mes},t]}({\hat \Delta})))\eqdef
%F(\hat\Delta),
%\end{equation}
%where $\hat\Delta$ is the vector obtained by
%concatenating all the
%elements of the estimated uncertainty matrices $\Delta\hat{ A}$,
%$\Delta\hat{ B}$, $\Delta\hat{C}$ and $\Delta\hat{D}$, $ T_{\rm mes}$ is the period between
%two updates of the parameters by the extremum
%seeking algorithm, ${y_{e}}^{[t-\delta T_{\rm mes},t]}({\hat \Delta})$ denotes the tracking
%error
%signal during the time interval $[t-\delta T_{\rm mes},t]$ for $\Delta = \hat \Delta$
%and $\psi:\rr^p \rar \rr$ maps the signal obtained for a given
%uncertainty value $\hat \Delta\in\rr^p$ to a
%real variable which is a measure of the performance. In~\eqref{Q}, $\tilde F:\rr\rar \rr$ is such
%that
%$\tilde F(0)=0$, $\tilde F(s)>0$,
%for $s\neq 0$, and from $\psi$ and $\tilde F$ we define $F=F\circ\psi$, $F:\rr^p\rar\rr_{0+}$,
%where $F=0$ if and only if the measure of performance is $0$.

In order to ensure convergence of the MES
algorithm, $Q$ need to satisfy the following assumptions.
\begin{assumption} \label{robustmesass1}
The cost function $Q$ has a local minimum at
$\hat\Delta^{*}=\Delta$.
\end{assumption}
\begin{assumption} \label{robustmesass2}
The original parameter estimate vector $\hat\Delta$ is close
enough to the actual parameters vector $\Delta$.
\end{assumption}
\begin{assumption} \label{robustmesass3}
The cost function is analytic and its variation with respect to
the uncertain variables is bounded in the neighborhood of
$\Delta^{*}$, i.e., there exists $\xi_{2}>0$, s.t.
$\|\frac{\partial{Q}}{\partial
\Delta}({\tilde{\Delta}})\|\leq\xi_{2}$ for all
$\tilde{\Delta}\in\mathcal{V}(\Delta^{*})$, where
$\mathcal{V}(\Delta^{*})$ denotes a compact neighborhood of
$\Delta^{*}$.
\end{assumption}
\begin{remark} Assumption \ref{robustmesass1} simply means that we can consider that $Q$ has
at least a local minimum at the true values of the uncertain
parameters.
\end{remark}
\begin{remark}
 Assumption \ref{robustmesass2} indicates
that our result will be of local nature, meaning that our analysis
holds in a small neighborhood of the actual values of the
parameters.
\end{remark}
\begin{remark}
We wrote the cost function (\ref{Q}) as function of the  tracking
error (\ref{tracking_error}), however, the case of regulation or
stabilization can be directly deduced from this formulation by
replacing the time-varying reference with a constant reference or
an equilibrium point.
\end{remark}

Under Assumptions \ref{robustmesass1}, \ref{robustmesass2}, and
\ref{robustmesass3}, it has been shown (e.g. \cite{AK02,Nes09}),
that the  MES
\begin{equation}
\begin{array}{l}
\dot{z}_{i}=a_{i}\mathrm{sin}(\omega_{i}t+\frac{\pi}{2})Q(\hat\Delta)\\
\hat\Delta_{i}=z_{i}+a_{i}\mathrm{sin}(\omega_{i}t-\frac{\pi}{2}),\;
i\in\{1,\ldots ,N_p\} \label{mesgenericbrake}
\end{array}
\end{equation}
with $N_p \leq nn+nm+pn+pm$ is the number of uncertain elements,
$\omega_{i}\neq\omega_{j},\;\omega_{i}+\omega_{j}\neq\omega_{k},\;i,j,k\in\{1,\ldots ,N_p
\}$, and $\omega_{i}>\omega^{*},\;\forall i\in\{1,\ldots ,N_p \}$,
with $\omega^{*}$ large enough, converges to the local minima of
$Q$.

The idea that we want to propose here (refer to \cite{B014} where we
introduced this concept of learning-based adaptive control for a
class of nonlinear systems), is that under Assumptions 1-6, we
can merge the MPC algorithm and a discrete-time version of
the MES algorithm to obtain an iterative learning MPC algorithm. We formalize this idea in the
following iterative
algorithm:\\\\
{\bf ALGORITHM I}
\begin{algorithmic}
\STATE {- Initialize $z_{i}(0)=0$, and the uncertainties' vector
estimate $\hat{\Delta}(0)=0$.}\\
\STATE {- Choose a threshold for the cost function minimization
$\epsilon_{Q}>0$.}\\
 \STATE{- Choose the
parameters, MPC sampling time $\delta T_{\rm mpc}>0$, and MES
sampling time $\delta T_{\rm mes}=N_E\delta T_{\rm mpc}$,
$N_E>0$.}\\
\STATE{- Choose the MES dither signals' amplitudes and
frequencies: $a_{i},\;\omega_{i},\;i=1,2\ldots, N_p $.}\\
 {{\bf WHILE}({\verb|true|})}

{\quad {\bf FOR}($\ell =1, \ell\leq N_E, \ell=\ell+1$)}
    \STATE{\quad \quad - Solve the MPC problem \bsube\label{eq:OCP_algo1}
  \beqar
\min_{ U(k)}&&\sum_{i=0}^{N-1}\|x(i|k)\|^2_{Q_{\rm M}}+\|{u(i|k)\|^2_{R_{\rm
M}}} \\ &&\quad+ \|x(N|k)\|^2_{P_{\rm M}}, \nonumber
  \\%+\|y(i|k)\|_2^{S_{\rm  M}} \\
   {\rm s.t.}
         && x(i+1|k)=(A+\Delta A)x(i|k) \\ && \qquad \qquad \qquad+(B+\Delta B)u(i|k), \nonumber
\\
         && y(i|k)=(C+\Delta C)x(i|k) \\ && \qquad \qquad \qquad +(D+\Delta D)u(i|k), \nonumber
\\
        && \xmin\leq x(i|k) \leq \xmax,\  i\in\zz_{[1,N_c]},
 \\
       &&\umin\leq  u(i|k) \leq \umax,\  i\in\zz_{[0,N_{cu}-1]},
\\
       &&\ymin\leq  y(i|k) \leq \ymax,\  i\in\zz_{[1,N_c]},
 \\
        &&  u(i|k)=K_f x(i|k),\  i\in\zz_{[N_u,N-1]},
 \\
        && x(0|k) = x(k),
  \eeqar\esube }
  \STATE {\quad \quad - Update $k=k+1$}.
      \\{\quad \bf End }

    {\bf \quad  IF}
  {$Q > \epsilon_{Q}$}

  \STATE{\quad \quad - Evaluate the MES cost function $Q(\hat{\Delta})$}
  \STATE{\quad  \quad - Evaluate the new value of the uncertainties  $\hat\Delta$:\\
\begin{equation}\label{MES_algo1}
\begin{array}{l}
{z}_{i}(h+1)=z_{i}(h)+a_{i}\delta T_{\rm mes} \sin(\omega_{i}h\delta T_{\rm
mes}+\frac{\pi}{2})Q(\hat\Delta)\\
\hat\Delta_{i}(h+1)=z_{i}(h+1)+a_{i}\sin(\omega_{i}h\delta
T_{\rm mes}-\frac{\pi}{2}),\\ i\in\{1,\ldots ,N_p\}
\end{array}
\end{equation}
  }\STATE {\quad \quad - Update $h=h+1$}.
    \\{\quad \bf End }
    \\{\quad Reset $\ell=0$}
    \\{\bf End }
\end{algorithmic}

\vspace{-0.2cm}
\subsection{Stability discussion}
As mentioned, in this preliminary work we aim at presenting an
algorithm that merges model-based MPC and model-free MES learning,
to obtain an iterative learning MPC algorithm, with encouraging
numerical numerical results (see, e.g., the case study in Section
\ref{section3}).  A rigorous stability analysis of the combined
algorithm is out of the scope of this work, but  we want to sketch
below an approach to analyze the stability of Algorithm I. We
propose here to follow the analysis presented in \cite{B014}, for
the case of learning-based adaptive control for some class of
nonlinear systems. By Assumptions 1 -3, the model structural
uncertainties $\Delta A, \Delta B, \Delta C$, and $\Delta D$ are
bounded, the uncertain model matrices $A+\Delta A$, $B+\Delta B$,
$C+\Delta C$, and $D+\Delta D$ are elements of convex sets
$\mathcal{K}_{a}$, $\mathcal{K}_{b}$, $\mathcal{K}_{c}$, and
$\mathcal{K}_{d}$, and that the MPC problem (\ref{eq:OCP_algo1})
is well-posed. Based on this, the approach for proving stability
is based on establishing a boundedness of the tracking error norm
$\|y_{e}\|$ with the upper-bound being function of the
uncertainties estimation error norm $\|\hat{\Delta}-\Delta\|$. One
effective way to characterize such a bound is to use an integral
Input-to-State Stability (iISS) (or ISS for time-invariant
problems) between the input $\|\hat{\Delta}-\Delta\|$ and the
augmented state $\|y_{e}\|$, see, e.g. \cite{B014}. If the iISS
(or ISS) property is obtained, by reducing the estimation error
$\|\hat{\Delta}-\Delta\|$ we also reduce the the tracking error
$\|y_{e}\|$, due to the iISS relation between the two signals.
Based on Assumptions 5, 6, and 7, we know, e.g. \cite{AK02,Nes09},
that the MES algorithm (\ref{MES_algo1}) converges to a local
minimum of the MES cost $Q$, which implies (based on Assumption
4), that the estimation error $\|\hat{\Delta}-\Delta\|$ is
reducing over the MES iterations. Thus, finally we conclude that
the MPC tracking (or regulation) performance is improved over the
iterations of Algorithm I. While obviously this discussion is not
a rigorous proof of stability of the iterative learning MPC, it
provides an interesting guideline to follow for analyzing the
controller stability observed during the test case presented next.
A more rigorous proof is currently under development and will be
presented in future works. \vspace{-0.3cm}
\section{DC servo-motor example}
\label{section3} The example studied here is about the angular
position control of a load connected by a flexible shaft to a
voltage actuated DC servo motor, see \cite{DBB13} and references therein. The states are the load
angle and angular rate, and the motor angle and angular rate, the
control input is the motor voltage, and the outputs are the load
angle and the torque acting on the flexible shaft. The model for
the system is
\begin{equation}\label{eq:dcmotor}
\begin{array}{l}
 \dot x_c(t)=\left[
 \begin{array}{cccc}
0  & 1   &   0  &    0\\
     -\frac{k_{l}}{\JJ_l}   &    -\frac{\beta_l}{\JJ_l}  &   \frac{k_{l}}{g\JJ_l}  &   0\\
     0        &     0         &    0           &      1 \\
     \frac{k_{l}}{g\JJ_m} & 0      &       -\frac{k_{l}}{g^2\JJ_m}  &
-\frac{\beta_m+R_A^{-1}K_m^2}{\JJ_m}
 \end{array}
\right]x_c(t)+\\
\hspace{+6cm}\left[
\begin{array}{c}
0 \\ 0 \\ 0 \\\frac{K_m}{R_A\JJ_m}
\end{array}
\right]u_c(t)\\
y_c(t)= \left[
\begin{array}{cccc}
1 & 0 & 0 & 0\\ k_l &  0 & -\frac{k_l}{g} & 0
\end{array}
 \right]x_c(t)
\end{array}
\end{equation}
where $x_c\in\rr^4$ is the state vector, $u_c\in\rr$ is the input
vector, and $y_c\in\rr^2$ is the output vector.
In~\eqref{eq:dcmotor} $R_A$[$\Omega$] is the armature resistance,
$K_m$[Nm/A] is the motor constant, $\JJ_l$[kgm$^2$],
$\beta_l$[Nms/rad], $k_l$[Nm/rad], are the inertia, friction and
stiffness of load and flexible shaft, $\JJ_m$[kgm$^2$],
$\beta_m$[Nms/rad], are the inertia and friction of the motor, and
$g$ is the gear ratio between motor and load. The nominal
numerical values used in the simulations are $R_A=10\Omega$,
$K_m=10$Nm/A, $\JJ_l=25$kgm$^2$, $\beta_l=25$Nms/rad,
$k_l=1.28\cdot 10^3$Nm/rad, $\JJ_m=0.5$kgm$^2$,
$\beta_m=0.1$Nms/rad. The system is subject to constraints on
motor voltage and shaft torque \bsube\label{eq:dcmotconstr}
 \beqar\label{eq:out-constraint}
 &   -78.5  \leq [y_c(t)]_2 \leq  78.5   ,\\
 &  -220 \leq u_c(t) \leq  220.
  \eeqar
  \esube

The control objective is to track a time varying load angle
position reference signal $r_l(t)$. Following the nominal MPC
presented in Section \ref{nominal_mpc}, the prediction model is
obtained by sampling \eqref{eq:dcmotor} with a period $\delta
T_{\rm mpc}=0.1s$, and the trajectory tracking problem is solved by
augmenting the model with the reference prediction model~\eqref{eq:refModel}, which in this case
represents a constant reference
   $$r_r(k+1)=r_r(k),$$
where $r_r\in\rr$,   and with an incremental formulation of the
control input, such that
   $$u_c(k+1)=u_c(k)+\Delta v(k)$$
Next, the MPC cost function is chosen as
  \beq\label{eq:cost}
 \sum_{i=1}^N\|[y(i|k)]_1-r_l(i|k)\|^2_{Q_y}+\|\Delta v(i|k)\|^2_{R_v}+\rho
\sigma^2,
 \eeq
where $Q_y= 10^3 $ and $R_v=0.05$,  prediction, constraints, and
control horizons are $N=20$, $N_{c}=N_{cu}=N_u=4$, and $K_f=0$
in~\eqref{eq:OCP-tControl}. In this case study output
constraints~\eqref{eq:out-constraint} are considered as soft
constraints, which may be (briefly) violated due to the prediction
model not being equal to the actual model. Thus, in
\eqref{eq:cost} we add the term $\rho\sigma^{2}$, where $\rho>0$
is a (large) cost weight, and $\sigma$ is an additional variable
used to model the (maximum) constraint violation of the softened
constraints.

Thus, the MPC problem~\eqref{eq:OCP} results in a family of
quadratic programs parameterized by the current state $x(k)$
in~\eqref{eq:OCP-initX}, with $n_q=16$ constraints, and $n_u=4$
variables. In the subsequent simulations, we consider initial
state  $x(0)=[0\ 0\ 0\ 0]'$ and reference
$r_l(t)=4.5\sin(\frac{2\pi}{T_{ref}}t)$, $T_{ref}=20\pi\;sec$.

\begin{figure}
\includegraphics[scale=0.2]{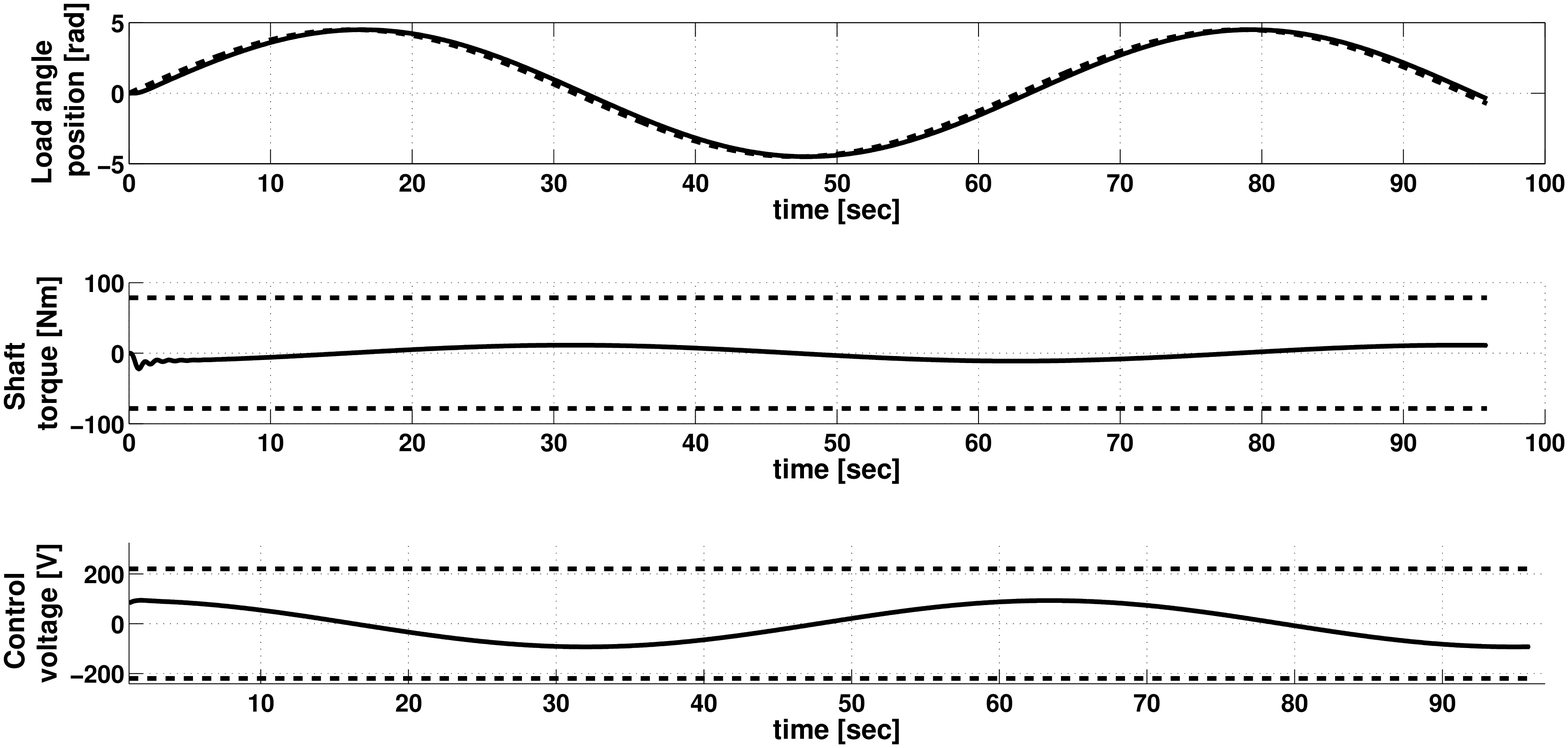}
\caption{Outputs and input signals in the nominal case (reference
trajectory and constraints limits in dashed-line, obtained signals
in solid-line)}\label{fig0}
\end{figure}
\begin{figure}
\includegraphics[scale=0.2]{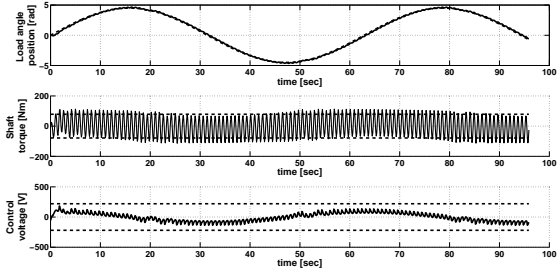}
\caption{Outputs and input signals in the uncertain case with
nominal MPC control (reference trajectory and constraints limits
in dashed-line, obtained signals in
solid-line)}\label{fig0_uncertain}
\end{figure}

First, to have a base-line
performance, we solve the nominal MPC problem, i.e. without model
uncertainties. We report the corresponding results on Figure
\ref{fig0}, where it is clear that the desired load angular
position is precisely tracked, without violating the shaft torque
and the control voltage constraints. Next, we introduce the
parametric model uncertainty $\delta\beta_{l}=-70\;[Nm s/rad]$.
Note that we purposely introduced a very large model uncertainty,
i.e. more than $100\%$ of the nominal value, to show clearly the
bad effect of this uncertainty on the nominal MPC algorithm and to
subsequently test the iterative learning MPC algorithm on a challenging case.
We first apply the nominal MPC controller to the uncertain model,
we show the obtained performance on Figure  \ref{fig0_uncertain},
where it is clear that the nominal performance is lost, since the
second output, i.e. the shaft torque, is oscillating and is
violating its upper and lower limits. Furthermore, this
oscillations are also present on the control voltage signal.

Now, we apply the  iterative learning MPC Algorithm I, where we
set $\delta T_{\rm mes}=1.5T_{ref}$. We choose the MES learning
cost function as \beqar \no
Q&=&\sum_{i=0}^{N_E-1}\|[y_{e}(t-i\delta
T_{mpc})]_{1}\|^2+\|[\dot{y}_{e}(t-i\delta
T_{mpc})]_{1}\|^2\\
 &&\qquad +\|[y_{e}(t-i\delta T_{mpc})]_{2}\|^2,\no
\eeqar
i.e. the norm of the error in the load angular position and
velocity, plus the norm of the error on the shaft torque. To learn
the uncertain parameter $\beta_{l}$, we apply the algorithm
(\ref{MES_algo1}), as
\begin{equation}\label{beta_l_learning}
\begin{array}{l}
{z}_{\beta_{l}}(k^{'}+1)=z_{\beta_{l}}(k^{'})+a_{\beta_{l}}\delta T_{mes} sin(\omega_{\beta_{l}}k^{'}\delta T_{mes}+\frac{\pi}{2})Q\\
\delta\hat{\beta}_{l}(k^{'}+1)=z_{\beta_{l}}(k^{'}+1)+a_{\beta_{l}}sin(\omega_{\beta_{l}}k^{'}\delta
T_{mes}-\frac{\pi}{2}),
\end{array}
\end{equation}
with $a_{\beta_{l}}=10^{-6},\;\omega_{\beta_{l}}=0.7\;rad/s$. We
select $\omega_{\beta_{l}}$ to be higher than the desired
frequency of the closed-loop (around $0.1\;rad/s$), to ensure
convergence of the MES algorithm, since the ES algorithms
convergence proofs are based on averaging theory, which assumes
high dither frequencies, e.g. \cite{AK02,Nes09}. We chose a small
value of the dither signal amplitude, since we noticed that the
MES cost function has large values due to the large simulated
uncertainty, so to keep the search excursion amplitude small, and
converge to a precise value of the uncertainty $\delta\beta_{l}$,
i.e. to keep $a_{\beta_{l}}Q$ small, we choose small
$a_{\beta_{l}}$ (for further explanations on how to tune MES
algorithms please refer to \cite{TNM08}). We also set the MES cost
function threshold $\epsilon_{Q}$ to $1.5 Q_{nominal}$, where
$Q_{nominal}$ is the value of the MES cost function obtained in
the nominal-model case with the nominal MPC, i.e. the base-line
ideal case. In other words, we decide to stop searching for the
best estimation of the value of the uncertainty when the uncertain
MES cost function, i.e., the value of $Q$ when applying the iterative learning MPC
algorithm to the uncertain model, is less or equal to $1.5$ of the
MES cost function in the case without model uncertainties, which represents the best achievable
MES cost
function.

\begin{figure}
\includegraphics[scale=0.2]{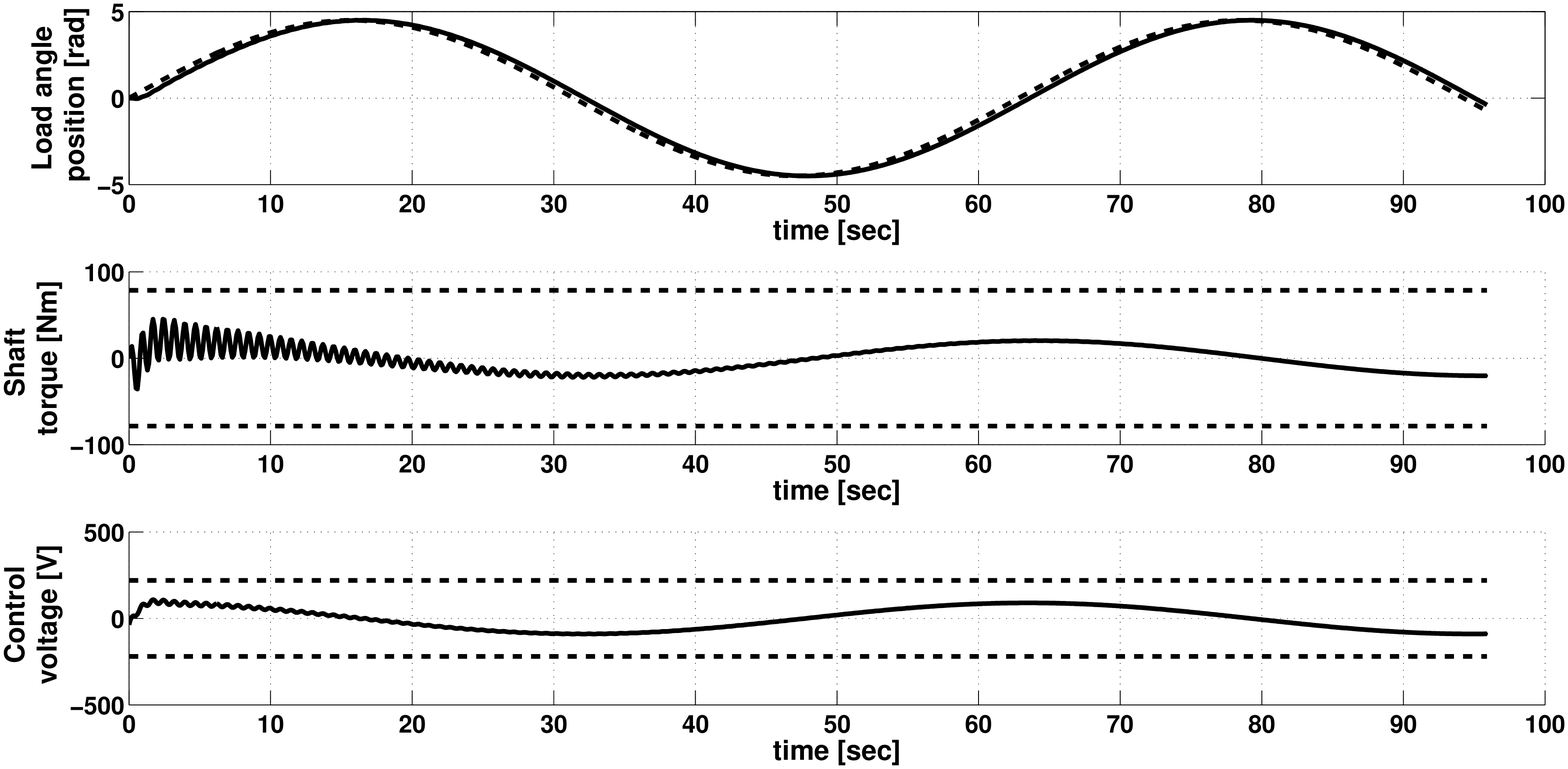}
\caption{Outputs and input signals in the uncertain case with
iterative learning MPC (reference trajectory and constraints
limits in dashed-line, obtained signals in
solid-line)}\label{fig1}
\end{figure}

\begin{figure}\center
\includegraphics[scale=0.3]{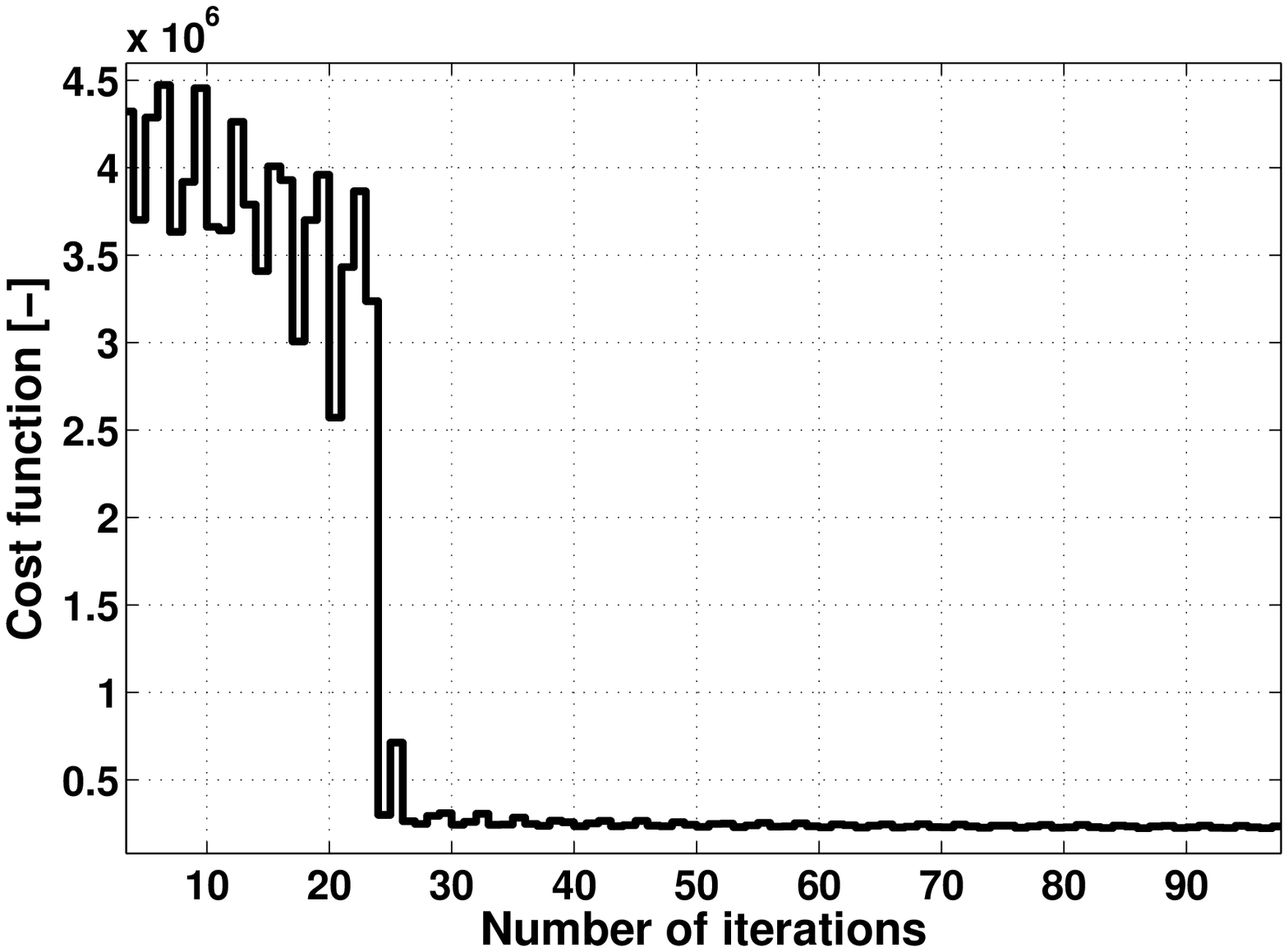}
\caption{MES cost function evolution over the learning
iterations}\label{fig2}
\end{figure}

\begin{figure}\center
\includegraphics[scale=0.3]{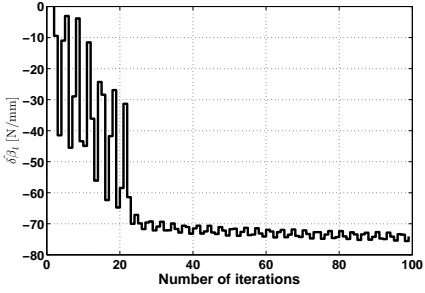}
\caption{Uncertain parameter learning evolution over the learning
iterations}\label{fig3}
\end{figure}

The obtained results of the iterative learning MPC algorithm are
reported on Figures \ref{fig1}, \ref{fig2} and \ref{fig3}. First,
note on Figure  \ref{fig2}, that the uncertain cost function
initial value (at the first iteration of the MES learning) is very
high, about $4.5\cdot 10^6$, which is about $50$ times the value
of the nominal base-line MES cost function value. We see on Figure
\ref{fig2} that this cost function decreases as expected along the
MES learning iterations to reach a small value after about $25$
iterations. This corresponds to the required number of iterations
to learn the actual value of the uncertain parameter as shown on
Figure  \ref{fig3}. Eventually, after the convergence of the
learning algorithm, we see on Figure  \ref{fig1}, that the nominal
base-line performances of the MPC are recovered and that the
output track the desired reference with smooth signals and without
violating the desired constraints.

\begin{figure}\center
\includegraphics[scale=0.2]{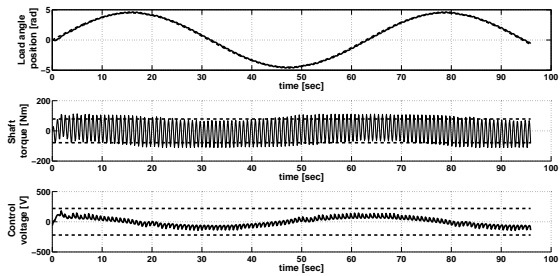}
\caption{Outputs and input signals in the uncertain case with
nominal MPC control (reference trajectory and constraints limits
in dashed-line, obtained signals in
solid-line)}\label{fig4_uncertain}
\end{figure}

We also tested the case of
multiple uncertainties. We assumed the two uncertainties
$\delta{\beta}_{l}=-70,\;[Nm s/rad]$,
$\delta{J}_{l}=-0.2,\;[kgm^{2}]$. We first show on Figure
\ref{fig4_uncertain} the performance of the nominal MPC when
applied to this uncertain model. It is clear that the nominal MPC
cannot cope with the uncertainties effect on the system's
performance, since the shaft torque, is oscillating and is
violating its limits. The control voltage signal experiences
oscillations, as well. Next, we apply the iterative learning-based
MPC, where $\hat{\delta\beta}_{l}$ is learned using
(\ref{beta_l_learning}), and $\hat{\delta J}_{l}$ is learned using
the ES equations
\begin{equation}\label{j_l_learning}
\begin{array}{l}
{z}_{J_{l}}(k^{'}+1)=z_{J_{l}}(k^{'})+a_{J_{l}}\delta T_{mes} sin(\omega_{J_{l}}k^{'}\delta T_{mes}+\frac{\pi}{2})Q\\
\delta\hat{J}_{l}(k^{'}+1)=z_{J_{l}}(k^{'}+1)+a_{J_{l}}sin(\omega_{J_{l}}k^{'}\delta
T_{mes}-\frac{\pi}{2}),
\end{array}
\end{equation}
with $a_{J_{l}}=10^{-8}$, and $\omega_{J_{l}}=0.8\;rad/s$. Note
that we choose a smaller dither amplitude for the $\delta J_{l}$
estimation, since the value of the uncertainty on $J_{l}$ is
smaller, so we need a smaller dither signal amplitude for the
search of the uncertain value.

\begin{figure}\center
\includegraphics[scale=0.2]{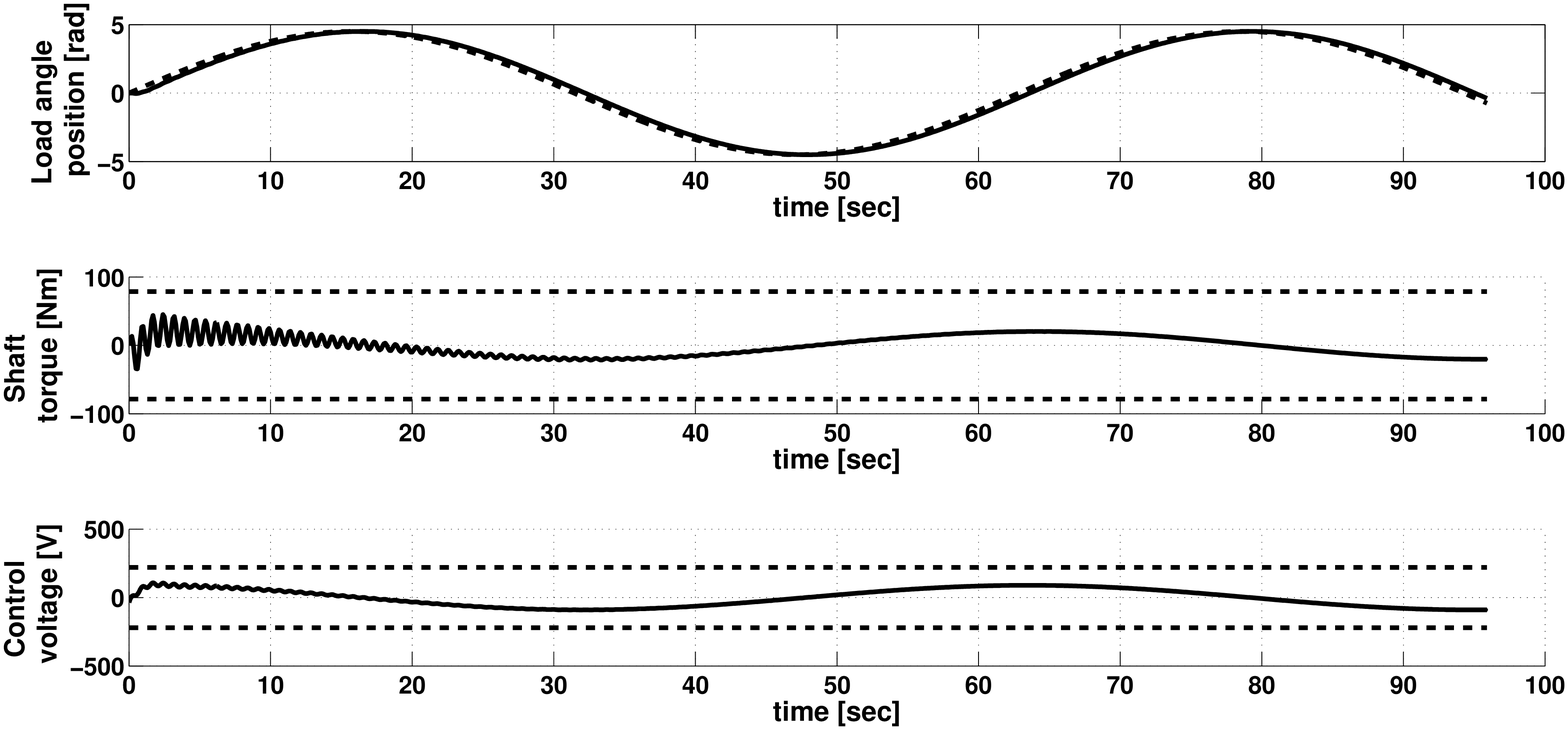}
\caption{Outputs and input signals in the uncertain case with
iterative learning MPC (reference trajectory and constraints
limits in dashed-line, obtained signals in
solid-line)}\label{fig4}
\end{figure}

\begin{figure}\center
\includegraphics[scale=0.3]{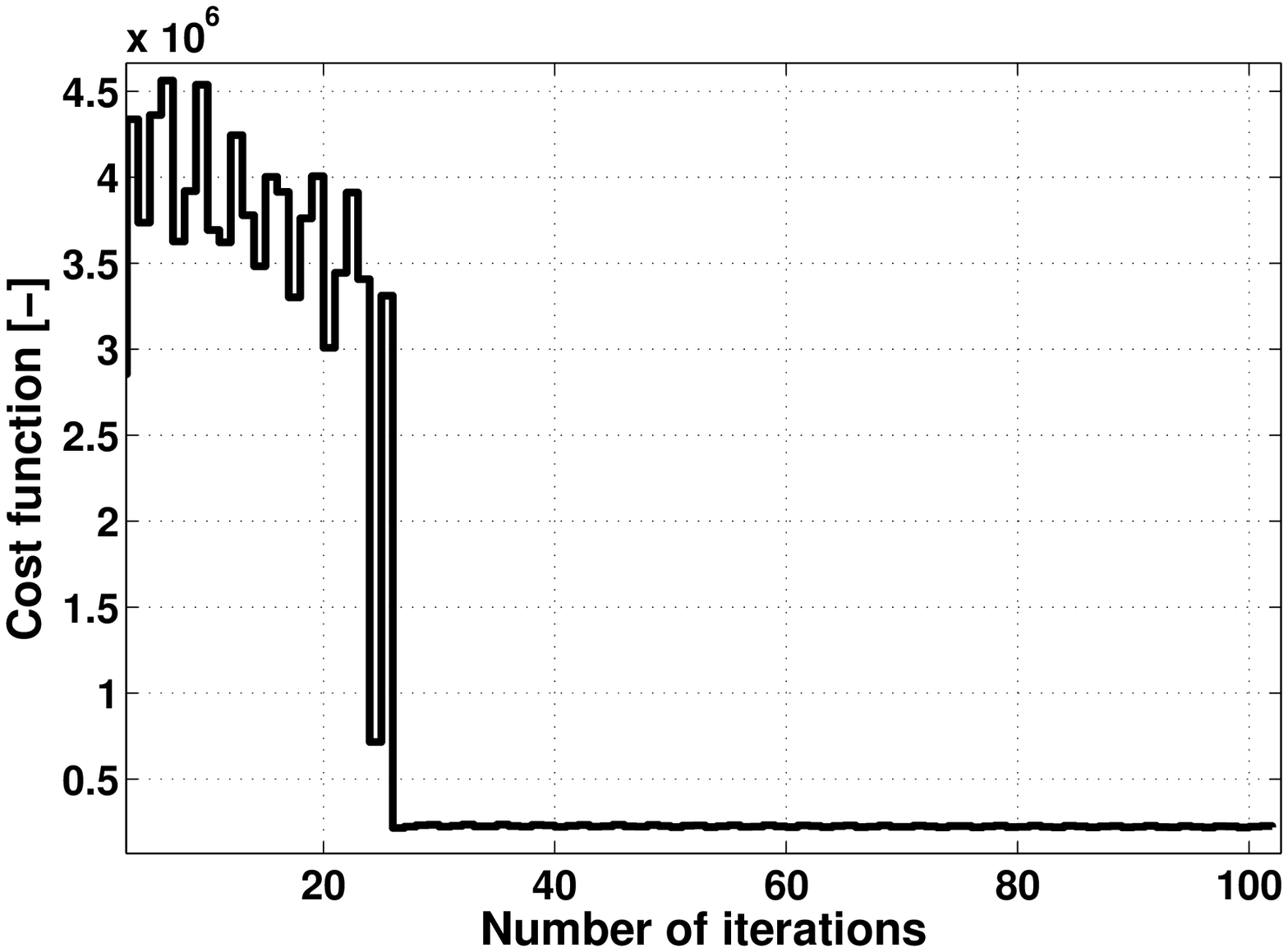}
\caption{MES cost function evolution over the learning
iterations}\label{fig5}
\end{figure}

\begin{figure}\center
\includegraphics[scale=0.3]{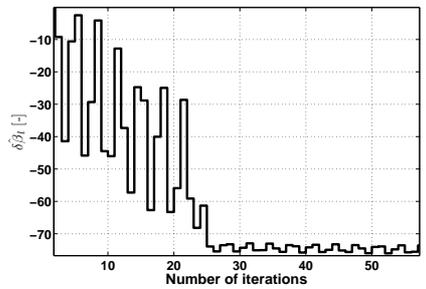}
\caption{Uncertain parameter $\delta\beta_{l}$ learning evolution
over the learning iterations}\label{fig6}
\end{figure}

\begin{figure}\center
\includegraphics[scale=0.3]{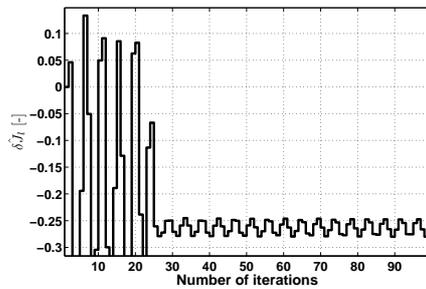}
\caption{Uncertain parameter $\delta J_{l}$ learning evolution
over the learning iterations}\label{fig7}
\end{figure}

The results of the iterative learning MPC algorithm in this case
are reported on Figure s \ref{fig4}, \ref{fig5}, \ref{fig6} and
\ref{fig7}. The learning cost function shown on Figure \ref{fig5},
is clearly decreasing and stabilizes after about $30$ iterations.
The uncertainties are learned and the overall tracking performance
is recovered, as shown on Figures \ref{fig6}, \ref{fig7} and
\ref{fig4}, respectively. We notice here that the estimation of
the uncertainties has some small residual error, this estimation
error can be improved by either fine tuning the MES dither
signals' amplitudes, e.g., by using a time-varying dither
amplitude \cite{MMB09}, or by choosing other type of MES
algorithms with larger domain of attraction, e.g.
\cite{TNM06,MTNM11}.

\section{Conclusion}\label{section4}
In this paper, we have reported some preliminary results about an
MES-based adaptive MPC algorithm. We have argued that it is
possible to merge together a model-based linear MPC algorithm with
a model-free MES algorithm to iteratively learn structural model
uncertainties and thus improve the overall performance of the MPC
controller. We have discussed a possible direction to analyze the
stability of such algorithms. However, a more rigorous analysis of
the stability and convergence of the proposed algorithm is under
development, and will be presented in our future reports. We have
reported encouraging numerical results obtained on a mechatronics
example, namely, a DC servo-motor control example. Future
investigations will focus on improving the convergence rate of the
iterative  learning MPC algorithm, by using different ES
algorithms with semi-global convergence properties, e.g.
\cite{TNM06,MTNM11}, and on extending this work to different types
of model-free learning algorithms, e.g. reinforcement learning
algorithms, and comparing the learning algorithms in terms of
their convergence rate and achievable optimal performances.

\end{document}